\documentclass[sigconf, language=french, language=english]{acmart}
\AtBeginDocument{%
  }

\setcopyright{none}
\copyrightyear{2024}
\acmDOI{}
\acmISBN{}
\acmConference[40ème conférence sur la gestion des données (BDA)]{BDA}{October 21-24, 2024}{Orléans, France}
\usepackage{listings}
\usepackage{xcolor}
\usepackage{multirow}
\usepackage{stfloats}

\begin{document}

%%
%% style listing (code block)
\lstdefinelanguage{SPARQL}{
  basicstyle=\small\ttfamily,
  columns=fullflexible,
  breaklines=true,
  sensitive=true,
  % --------------------------
  tabsize = 2,
  showstringspaces=false,
  morecomment=[l][\color{green}]{\#},       % comments
  morecomment=[n][\color{red}]{<}{>}, % uris
  morestring=[b][\color{orange}]{\"},  % strings
  % -------------------------- variables
  keywordsprefix=?,
  classoffset=0,
  keywordstyle=\color{darkgray},
  morekeywords={},
  % -------------------------- prefixes
  classoffset=1,
  keywordstyle=\color{purple},
  morekeywords={rdf,vers},
  % -------------------------- keywords
  classoffset=2,
  keywordstyle=\color{blue},
  morekeywords={
    SELECT,CONSTRUCT,DESCRIBE,ASK,WHERE,FROM,NAMED,PREFIX,BASE,OPTIONAL,
    FILTER,GRAPH,LIMIT,OFFSET,SERVICE,UNION,EXISTS,NOT,BINDINGS,MINUS,a
  }
}

\title{ConVer-G: Concurrent versioning of knowledge graphs}
\translatedtitle{french}{ConVer-G: Versionnement concurrent de graphes de connaissances}

\author{Jey Puget Gil}
\email{jey.puget-gil@liris.cnrs.fr}
\orcid{0009-0006-6198-7488}
\affiliation{%
  \institution{Université Claude Bernard, LIRIS, UMR-CNRS 5205}
  \city{Villeurbanne}
  \country{FRANCE}
}

\author{Emmanuel Coquery}
\email{emmanuel.coquery@univ-lyon1.fr}
\orcid{}
\affiliation{%
  \institution{Université Claude Bernard, LIRIS, UMR-CNRS 5205}
  \city{Villeurbanne}
  \country{FRANCE}
}

\author{John Samuel}
\email{john.samuel@cpe.fr}
\orcid{0000-0001-8721-7007}
\affiliation{%
  \institution{Université de Lyon, CPE Lyon, LIRIS, UMR-CNRS 5205}
  \city{Villeurbanne}
  \country{FRANCE}
}

\author{Gilles Gesquière}
\email{gilles.gesquiere@univ-lyon2.fr}
\orcid{0000-0001-7088-1067}
\affiliation{%
  \institution{Université de Lyon, Université Lumière Lyon 2, LIRIS, UMR-CNRS 5205}
  \city{Villeurbanne}
  \country{FRANCE}
}

\renewcommand{\shortauthors}{Puget Gil et al.}

%       RDF macros      %
\newcommand{\term}[1]{
    t_{#1}
}
\newcommand{\subject}{
    \term{s}
}
\newcommand{\predicate}{
    \term{p}
}
\newcommand{\object}{
    \term{o}
}
\newcommand{\namedgraph}{
    \term{g}
}
\newcommand{\nnamedgraph}[1]{
    \term{g#1}
}
\newcommand{\rdftriple}{
    (\subject, \predicate, \object)
}
\newcommand{\rdfquad}{
    (\subject, \predicate, \object, \namedgraph)
}

\begin{abstract}
    The multiplication of platforms offering open data has facilitated access to information that can be used for research, innovation, and decision-making. Providing transparency and availability, open data is regularly updated, allowing us to observe their evolution over time.

    We are particularly interested in the evolution of urban data that allows stakeholders to better understand dynamics and propose solutions to improve the quality of life of citizens. In this context, we are interested in the management of evolving data, especially urban data and the ability to query these data across the available versions. In order to have the ability to understand our urban heritage and propose new scenarios, we must be able to search for knowledge through concurrent versions of urban knowledge graphs.

    In this work, we present the ConVer-G (Concurrent Versioning of knowledge Graphs) system for storage and querying through multiple concurrent versions of graphs.
\end{abstract}

\begin{translatedabstract}{french}
    La multiplication de plateformes offrant des données ouvertes a permis de faciliter l'accès à des informations pouvant être utilisées pour la recherche, l'innovation et la prise de décision. Apportant transparence et disponibilité, les données ouvertes sont mises à jour régulièrement, nous pouvons alors observer leurs évolutions à travers le temps.

    Nous nous intéressons plus particulièrement à l'évolution des données urbaines qui permet à des acteurs de mieux comprendre les dynamiques et de proposer des solutions pour améliorer la qualité de vie des citoyens. C'est dans ce contexte que nous nous intéressons à la gestion des données urbaines et à la nécessité de pouvoir interroger ces données à travers les versions disponibles. Afin d'avoir la capacité de comprendre notre héritage urbain et de proposer de nouveaux scénarios, nous devons être en mesure de chercher de la connaissance à travers des versions concurrentes des graphes de connaissances urbaines.

    À travers ces travaux, nous présentons le système ConVer-G (Versionnement Concurrent de Graphes de Connaissances) permettant le stockage et l'interrogation de multiples versions concurrentes de graphes.
\end{translatedabstract}

\begin{CCSXML}
  <ccs2012>
  <concept>
  <concept_id>10002951.10003227.10003236.10003237</concept_id>
  <concept_desc>Information systems~Geographic information systems</concept_desc>
  <concept_significance>100</concept_significance>
  </concept>
  <concept>
  <concept_id>10002951.10003260.10003309.10003315.10003314</concept_id>
  <concept_desc>Information systems~Resource Description Framework (RDF)</concept_desc>
  <concept_significance>500</concept_significance>
  </concept>
  </ccs2012>
\end{CCSXML}

\ccsdesc[100]{Information systems~Geographic information systems}
\ccsdesc[500]{Information systems~Resource Description Framework (RDF)}

\keywords{RDF, versioning, concurrent versioning, graph, urban data, query}
\translatedkeywords{french}{RDF, versionnement, versionnement concurrent, graphe, données urbaines, requête}
%   \includegraphics[width=\textwidth]{assets/evolution-urbaine.pdf}
%   \caption{Concurrent versioning of urban data in a neighborhood}
%   \Description{Versionnement concurrent de données urbaines dans un quartier.}

\received{June 3, 2024}
\received[accepted]{July 12, 2024}
\received[revised]{\today}

\maketitle

\section{Introduction and Motivation}

Urban environments, with their complex and ever-changing nature, particularly benefit from open data. Effective urban management requires a deep understanding of various dynamic elements, such as population growth, infrastructure development, and transport frequentations. Open data allows urban planners, policymakers, and researchers to gain insights into these dynamics, fostering better planning and decision-making processes.

Transforming 3D geospatial urban data to a knowledge graph \cite{vinasco2021towards} enables the representation of complex urban environments in a structured and semantically rich manner. Concurrent viewpoints of urban evolution can be captured through the versioning of these knowledge graphs. Moreover, this versioning allows the representation of different states of the urban environment over time, providing a comprehensive view of the city's evolution. 

The versioning of scientific data also represents a certain interest in order to be able to exploit them\cite{klump2021versioning}. Indeed, it’s crucial to be able to reproduce experiments and analyses. If the data changes without versioning, it becomes difficult to reproduce the results at a later date. Versioning can help to protect the integrity of the data. If an error is introduced into the data, having versioned backups allows the data to be restored to a previous correct state. Versioning allows them to easily switch between different versions of the data for these experiments.

In this paper, we present a knowledge graph loader component (QuaDer\footnote{Quads loaDer}) and a versioned graph query system (QuaQue\footnote{Quads Query}). We have implemented demonstration scenarios to show the feasibility of our approach. The first scenario uses a real dataset representing a district in Villeurbanne called "Gratte-Ciel" to query the knowledge graph using versioned SPARQL queries. The second scenario uses a synthetic dataset to query a knowledge graph with a large number of versions. We aim to show that our system can query a large number of versions simultaneously and that the query response time remains acceptable.

\section{Related work}
As mentioned by Bayoudhi et al. \cite{bayoudhi2020survey}, versioning is a critical aspect of management in a project. It allows developers to track changes to the elements, manage different versions, and collaborate with other team members.

\subsection{Code versioning}
Version control systems \cite{ZOLKIFLI2018408} such as Git, Mercurial, Subversion, CVS, and Bazaar are commonly used for managing the evolution of source code. However, these systems are not specifically designed for data versioning, especially for structured data and have certain limitations. Apart from exceptions, like Git LFS, they don't provide adequate support for large files and data query. Therefore, they may not be the ideal choice for managing data versioning in a project.

\subsection{Dataset versioning}
\subsubsection{Snapshot-based versioning}
There are several versioning tools available for managing data in a project. Some popular Git-based tools (snapshot-based versioning tools) like Qri\footnote{\url{https://qri.dev/}}, QuitStore\cite{arndt-2020-dissertation}, GeoGig\footnote{\url{https://geogig.org/}}, and UrbanCo2Fab\cite{samuel2018urbanco2fab} represent versions as snapshots. These tools provide version control capabilities specifically designed for data versioning. However, one limitation of these tools is that in order to query a version, it needs to be checked out. This limits the capacity of such systems to answer queries over multiple versions at the same time. Despite this limitation, database versioning tools provide valuable functionality for managing data in a project.

On the other hand, there are also machine learning tools like DVC\cite{ruf2021demystifying}, DagsHub\footnote{\url{https://dagshub.com/}} and Weights and Biases\footnote{\url{https://wandb.ai/}} that offer advanced features for managing data versions. These tools are particularly useful for structured data and provide functionalities such as tracking changes, managing different versions, and collaborating with team members. 

\subsubsection{Interval-based versioning}
Another approach to data versioning involves using temporal tables or bitemporal tables \cite{10.1145/2380776.2380786} to represent the validity of data over time. These tables store the start and end timestamps for each version of the data, allowing for querying the data at specific points in time or in a specific time interval. However, this approach has limitations, in particular the lack of support for easily identifying and querying concurrent versions.

\subsubsection{Delta-based versioning}
Delta-based versioning systems like Delta Lake, Apache Hudi and OSTRICH (Offset-enabled TRIple store for CHangesets) \cite{taelman2019triple} provide advanced features for managing data versions. These systems store the changes made to the data as deltas, allowing for efficient querying of the data at different versions. They also provide functionalities such as time travel, which allows users to query the data at specific points in time.

\section{Use case and contributions}
\subsection{The problem}
To avoid data redundancy in a quickly changing world, where each change varies little from one state to another, we must condense the stored evolution. This is often necessary for urban datasets at larger scales. The problem we aim to solve with QuaQue is the efficient querying of concurrent versioning of quad data. Traditional quad stores are designed to efficiently store and retrieve static RDF data, but they struggle when it comes to handling concurrent versioning of data and more specifically multiple versions at the same time. Concurrent versioning of quad data refers to the ability to store and query multiple versions of the same dataset at the same time. This is particularly useful in the context of urban data, where multiple versions of the same dataset can represent different points of view or different states of the data over time.

In the RDF model, a triple is a statement that consists of a subject and an object connected by a predicate. A quad is an extension of the triple that includes a fourth element, a named graph. A named graph can be represented by a set of triples sharing the same graph name. In order to version a named graph, we associate a version with this set of quads. In SPARQL, the \textbf{graph} operator is used to specify a graph name. SPARQL does not provide a way to specify a version of a named graph. We do not want to create a new operator to specify a version of a named graph because it would break the compatibility with existing SPARQL queries. Instead we slightly change the semantics of the graph operator to specify a versioned named graph. We created URI identifiers for each versioned named graphs and store them as metadata. The versioned graph identifier allows us to identify the (graph name, version) pair of a versioned named graph. By associating a variable with it, the graph operator allows us to query the set of versioned graphs.

\subsection{RDF Context Representation}
\subsubsection{Dataset representation}
Consider our RDF context containing data as well as its associated metadata. We define a dataset as a set of versioned named graphs and metadata. The metadata is stored in the default graph. $d = (M, \left\{((v_1, \term{g1}), G_{1,1}), ..., ((v_m, \term{gn}), G_{m,n})\right\})$ is a dataset where:
\begin{itemize}
    \item we call a versioned graph $G_{m,n}$ a finite subset of triples that occurs at the version $v_m$ of the name graph $\term{gn}$,
    \item $d$ represents a pair containing the metadata $M$ and a set of versioned graphs $\left\{((v_1, \term{g1}), G_{1,1}), ..., ((v_m, \term{gn}), G_{m,n})\right\}$,
    \item $d$ has $m$ versions and $(v_1, ..., v_m)$ is the set of versions,
    \item $d$ has $n$ named graphs and $(\nnamedgraph{1}, ..., \nnamedgraph{n})$ is the set of named graphs.
\end{itemize}

\begin{example}
    Let's assume that we have a dataset representing concurrent points of view about the height of some buildings. In the table \ref{dataset-2v-2ng} we present two versions with the following quads:
    
    \begin{table}[h]
        \caption{Dataset with 2 versions and 2 named graphs}
        \label{dataset-2v-2ng}
        \begin{tabular}{ |l|l|l|l|l| }
            \hline
            Version & Subject    & Predicate & Object & Named graph \\
            \hline
            \multirow{3}{1em}{1}
                    & ex:bldg\#1 & height    & 10.5   & ng:Gr-Lyon  \\
                    & ex:bldg\#2 & height    & 9.1    & ng:Gr-Lyon  \\
                    & ex:bldg\#1 & height    & 11     & ng:IGN      \\
            \hline
            \multirow{3}{1em}{2}
                    & ex:bldg\#1 & height    & 10.5   & ng:IGN      \\
                    & ex:bldg\#1 & height    & 10.5   & ng:Gr-Lyon  \\
                    & ex:bldg\#3 & height    & 15     & ng:Gr-Lyon  \\
            \hline
        \end{tabular}
    \end{table}
\end{example}

\paragraph{Flat model}
The flat model is a classic representation, it associates each quad with a version where it occurs.

$d = (G, \left\{(\term{g1}, G_1), ..., (\term{gn}, G_n)\right\})$ is a dataset where:
\begin{itemize}
    \item $G$ is the default graph storing the metadata
    \item $G_i$ is a named graph storing the quads
\end{itemize}

\begin{example}
    After some quads transformations, the Table \ref{dataset-4vng} represents \textit{the flat model} of the dataset versioning.
    
    \begin{table}
        \caption{Flat model of the dataset versioning}
        \label{dataset-4vng}
        \begin{tabular}{ |l|l|l|l| }
            \hline
            Subject    & Predicate     & Object     & Named graph \\
            \hline
            \multicolumn{4}{|c|}{Versioned quads}                 \\
            \hline
            ex:bldg\#1 & height        & 10.5       & vng:1       \\
            ex:bldg\#2 & height        & 9.1        & vng:1       \\
            \hline
            ex:bldg\#1 & height        & 11         & vng:2       \\
            \hline
            ex:bldg\#1 & height        & 10.5       & vng:3       \\
            ex:bldg\#3 & height        & 15         & vng:3       \\
            \hline
            ex:bldg\#1 & height        & 10.5       & vng:4       \\
            \hline
            \multicolumn{4}{|c|}{Metadata}                        \\
            \hline
            vng:1      & is-version-of & ng:Gr-Lyon &             \\
            vng:1      & is-in-version & v:1        &             \\
            vng:2      & is-version-of & ng:IGN     &             \\
            vng:2      & is-in-version & v:1        &             \\
            vng:3      & is-version-of & ng:Gr-Lyon &             \\
            vng:3      & is-in-version & v:2        &             \\
            vng:4      & is-version-of & ng:IGN     &             \\
            vng:4      & is-in-version & v:2        &             \\
            \hline
        \end{tabular}
    \end{table}
\end{example}
\section{QuaQue: A Queryable Versioned Quad Store}

\subsection{The query engine}
The problem we aim to solve is the efficient querying of concurrent versioning of quad data. Traditional quad stores are designed to efficiently store and retrieve static RDF data, but they struggle when it comes to handling concurrent versioning of data and more specifically multiple versions at the same time. Concurrent versioning of quad data refers to the ability to store and query multiple versions of a dataset. This is particularly useful in the context of urban data, where multiple versions of the same dataset can represent different points of view or different states of the data over time.

The challenge with concurrent versioning of quad data is that querying it requires considering the temporal, the concurrence and the structural aspects of the data. For example, we may want to retrieve all quads that were valid during a specific time interval or find the all versions where a quad matches certain criteria.

\subsection{Architecture}
The architecture of ConVer-G consists of three main components: the quads loader (QuaDer), the query engine (QuaQue), and the storage (PostgreSQL).

QuaDer is a loader component that is responsible for the versioning of quad data. It stores the different versions of the dataset in a condensed form, allowing for efficient storage of the data. QuaDer uses a relational database to store the different versions of the dataset. The system uses a bitstring representation to store the presence of quads in a set of versions. For each new version of the dataset, QuaDer adds a new bit to the bitstring representing the presence of the quad in that version. If the quad is present in the version, the bit is set to 1, otherwise it is set to 0.

\begin{example}
    The Table \ref{dataset-condensed} represents the condensed model of the dataset versioning. 
    
    \begin{table}
        \caption{Condensed model of the dataset versioning}
        \label{dataset-condensed}
        \begin{tabular}{ |c|c|c|c|c| }
            \hline
            \multicolumn{5}{|c|}{Versioned quads}                    \\
            \hline
            Subject    & Predicate & Object & Named graph & Versions \\
            \hline
            ex:bldg\#1 & height    & 11     & ng:IGN      & 10       \\
            ex:bldg\#1 & height    & 10.5   & ng:IGN      & 01       \\
            ex:bldg\#3 & height    & 15     & ng:Gr-Lyon  & 01       \\
            ex:bldg\#2 & height    & 9.1    & ng:Gr-Lyon  & 10       \\
            ex:bldg\#1 & height    & 10.5   & ng:Gr-Lyon  & 11       \\
            \hline
        \end{tabular}
    \end{table}
\end{example}

QuaQue is a queryable versioned quad store that is designed to query concurrent versioning of quad data. It is built on top of a Jena Fuseki edited version and uses a relational database to store and query multiple versions of the same dataset at the same time. QuaQue uses the previous relational database to query the different versions of the dataset at the same time.

\subsection{Demonstration scenarios}
We conducted a series of experiments to evaluate the ability of QuaQue in terms of query translation and scalability. We used two datasets and ran a set of queries against them:
\begin{itemize}
    \item a urban dataset with multiple versions representing the evolution of a city over time with concurrent versioning of quad data (the urban aspect of the dataset is not important for this paper). The dataset contains around 430000 quads reparitioned in 6 versioned graphs and 15000 metadata triples.
    \item a synthetic dataset created with BSBM \cite{bizer2009berlin}. This dataset is composed of more than 175000000 quads reparted in 1000 versioned graphs.
\end{itemize}

We ran two types of queries against the datasets. The first type of queries are non aggregative queries that retrieve all quads that match certain criteria. The second type of queries are more complex queries that involve aggregation.

\subsubsection{Non aggregative queries}
We ran a set of non aggregative queries against the datasets to evaluate the ability of QuaQue to retrieve all quads that match certain criteria. The queries involve using the bitstring representation to retrieve the quads that match certain criteria. To compute the existence of a quad in a version, we use the bitwise AND operation between all the bitstrings and the result is the bitstring representing the presence of the quad in the versions.

\begin{lstlisting}[language=SPARQL, caption={Retrieve all resources and versions where a triple matches certain criteria}]
SELECT ?version ?subj ?obj WHERE {
    GRAPH ?vng { ?subj rdf:type ?obj . }
    ?vng vers:is-in-version ?version .
}
\end{lstlisting}

This query is equivalent to the flat model. It retrieves all versioned quads and their associated version. It is useful to test the ability of QuaDer to insert correctly the versioned quads and the ability of QuaQue to retrieve them with their associated version.

\begin{lstlisting}[language=SPARQL, caption={Create the differences graph between two versioned graphs}]
SELECT ?subj ?pred ?obj WHERE {
{ SELECT ?subj ?pred ?obj WHERE {
    GRAPH <vng1> { ?subj ?pred ?obj . }
} } MINUS {
    SELECT ?subj ?pred ?obj WHERE {
    GRAPH <vng2> { ?subj ?pred ?obj . }
} } }
\end{lstlisting}

This query tests the ability of QuaQue to get the differences between two versioned graphs. It retrieves all quads that are present in the first versioned graph but not in the second versioned graph. It is useful to test the ability of QuaQue to compute the differences between two versioned graphs and compare the result with other versioned triple store like OSTRICH.

\subsubsection{Aggregative queries}
We ran a set of aggregative queries against the datasets to evaluate the ability of QuaQue to perform complex queries that involve aggregation. 

\begin{lstlisting}[language=SPARQL, caption={Find the maximum value of a resource by version}]
SELECT ?version MAX(?o) WHERE {
    GRAPH ?vng {
        ?s bsbm:v01/vocabulary/rating2 ?o .
    }
    ?vng vers:is-in-version ?version .
} GROUP BY ?version
\end{lstlisting}

\begin{lstlisting}[language=SPARQL, caption={Count the number of elements that match certain criteria by version}]
SELECT ?version COUNT(?subj) WHERE {
    GRAPH ?vng { ?subj rdf:type ?obj . }
    ?vng vers:is-in-version ?version .
} GROUP BY ?version
\end{lstlisting}

\begin{lstlisting}[language=SPARQL, caption={Count the number of elements that match certain criteria by graph}]
SELECT ?graph COUNT(?obj) WHERE {
    GRAPH ?vng { ?subj rdf:type ?obj . }
    ?vng vers:is-version-of ?graph .
} GROUP BY ?graph
\end{lstlisting}

\begin{lstlisting}[language=SPARQL, caption={Count the number of version that match certain criteria by graph}]
SELECT ?graph COUNT(DISTINCT ?version) WHERE {
    GRAPH ?vng { ?subj rdf:type "sensor" . }
    ?vng vers:is-in-version ?version ;
         vers:is-version-of ?graph .
} GROUP BY ?graph
\end{lstlisting}

These queries allow the analysis of the impact of the bit vector for the optimization of SPARQL queries. The work of Sarah Cohen \cite{cohen2006rewriting} allows us to rewrite queries with arbitrary aggregation functions using views. In our case, we then use the bit vector to optimize the computation of aggregation functions. For example, to compute the sum of values for a resource by version, we use the bitwise sum operation between the bit vectors representing the count values of the resource in each version (e.g., \textit{"What's the sum of the height of all buildings by version?"}).

\subsection{Discussion}
\subsubsection{Bitstring use}
The bitstring representation is a key feature of QuaDer that allows for efficient storage and retrieval of versioned quad data. By using a bitstring to represent the presence of quads in a set of versions, we use the GRAPH operator to get all graph's versions that satisties a basic graph pattern by performing a bitwise AND operation between the bitstring of all satisfied quads. This is due to the condensed representation. The more we are able to keep the condensed representation, the more we are able to optimize the query translation. In future work, we plan to investigate the use of the bitstring representation to optimize the translation of more complex queries, such as aggregation and sorting.

\subsubsection{Link with metadata}
The metadata is stored in the default graph. The metadata contains information about the dataset, such as the versions and named graphs. The metadata is used to associate a versioned named graph with its version (using the \textit{is-in-version} property) and named graph (using the \textit{is-version-of} property). We can also add metadata about the dataset, such as the creation date, the author, and the description. By storing the metadata in the default graph, we can easily access it using the basic graph pattern without GRAPH operator. This allows us to retrieve the metadata and use it to query the different versions of the dataset.

\subsubsection{Translation extension}
QuaQue partially implements the translation of SPARQL operators. We have only implemented the translation for the basic operators, such as SELECT, GRAPH, JOIN, GROUP BY and basic graph patterns. This choice was made to demonstrate the feasibility of the approach and to evaluate the capabilities of the condensed model on complex queries such as aggregation.

Covering all the SPARQL operators would greatly enhance the capabilities of QuaQue as a query engine. It would enable researchers and data scientists to perform complex queries and analysis on concurrent versioned quad data. By implementing the translation for all SPARQL operators, QuaQue would be able to handle a wide range of query types, including filtering, aggregation, sorting, and joining. This would allow users to express their queries in a familiar and expressive language, making it easier to explore and analyze the data. Additionally, extending the translation to all SPARQL operators would make QuaQue compatible with existing SPARQL query tools and libraries, enabling seamless integration with other data processing and analysis pipelines.

\subsubsection{Configuration of the versioning representation}
Extending the implementation of QuaDer to handle a target representation such as RDF-star, Java, relational database, or property graphs would allow to understand which storage engine would be the more efficient to manage condensed model. By incorporating support for these additional representations, QuaDer would offer researchers and data scientists the flexibility to choose the most suitable representation for their specific use cases and requirements.

Incorporating these target representations into QuaDer would not only enhance its versatility but also contribute to the scientific community by providing a comprehensive and extensible platform for concurrent versioning of quad data. Researchers and practitioners would have the freedom to choose the most appropriate representation based on their specific needs, leading to more accurate and efficient analysis of evolving datasets.

\subsubsection{RDF graphs annotation}
RDF graphs are a suitable choice for representing heterogeneous data. They provide a flexible and standardized way to express relationships between entities. Additionally, RDF-star\footnote{https://www.w3.org/2021/12/rdf-star.html} and Property Graphs \cite{angles2019rdf} offer advanced features for representing complex data structures. However, it's important to note that the representation of data in RDF graphs can be complex, which may introduce some limitations. We can use the RDF-star representation to annotate the RDF graphs with versions. This would allow us to query the different versions of the dataset.

\section{Conclusion}
In this chapter, we have explored the capabilities of QuaQue, which contains alternative SPARQL operators. QuaQue offers a new approach to querying evolving graphs by providing an alternative to classic SPARQL queries.

With QuaQue, we can achieve everything that a classic triple store can do. QuaQue modifies a set of operators and functions that allow us to query all versions of a graph. This means that we can retrieve historical data and analyze the evolution of the graph over \textit{transaction} time. This feature is particularly useful in scenarios where data changes frequently and we need to track the changes and perform analysis on different versions of the graph.

In future research, it would be interesting to integrate an hybrid approach that combines transaction versioning and existence versioning (where transaction versioning is used to track the changes in the database and existence versioning is used to track the changes of the entities in the real world). This would allow us to query the graph at a specific point in time and retrieve the historical data. This would be particularly useful in scenarios where we need to analyze the evolution of the graph over time and perform analysis on different versions of the graph.

\begin{acks}
This work, titled \emph{ConVer-G: Concurrent versioning of knowledge graphs}, is supported and funded by the IADoc@UDL (Université de Lyon, Universite Claude Bernard Lyon 1) and LIRIS UMR 5205. We would like to express our gratitude to the BD team and the members of the Virtual City Project\footnote{\url{https://projet.liris.cnrs.fr/vcity/}} for their invaluable advice and assistance.
\end{acks}

\bibliographystyle{ACM-Reference-Format}
\bibliography{bibliography.bib}

\appendix

\end{document}